\documentstyle[twoside,fleqn,espcrc2,epsf]{article}

\def\simgt{\,\rlap{\lower 3.5 pt\hbox{$\mathchar \sim$}}\raise 1pt \hbox {$>$}\,}
\def\simlt{\,\rlap{\lower 3.5 pt\hbox{$\mathchar \sim$}}\raise 1pt \hbox {$<$}\,}

\title{
Exploring QCD at small sea quark masses
with improved Wilson-type quarks\thanks{Talk presented by Y.~Namekawa}
}

\author{CP-PACS Collaboration : 
	Y.~Namekawa\rlap,\address{Institute of Physics, University of
        Tsukuba, Tsukuba 305-8571, Japan}
        S.~Aoki\rlap,$^{\rm a}$
        M.~Fukugita\rlap,\address{Institute for Cosmic Ray Research,
        University of Tokyo, Kashiwa 277-8582, Japan}
        K-I.~Ishikawa\rlap,$^{\rm a,}$\address{Center for Computational Physics,
        University of Tsukuba,Tsukuba 305-8577, Japan}
        N.~Ishizuka\rlap,$^{\rm a,c}$
        Y.~Iwasaki\rlap,$^{\rm a}$
        K.~Kanaya\rlap,$^{\rm a}$
        T.~Kaneko\rlap,\address{High Energy Accelerator Research 
        Organization(KEK), Tsukuba 305-0801, Japan}
        Y.~Kuramashi\rlap,$^{\rm d}$
	V.I.~Lesk\rlap,$^{\rm c}$
        M.~Okawa\rlap,\address{Department of Physics, Hiroshima University,
        Higashi-Hiroshima 739-8526, Japan}
	Y.~Taniguchi\rlap,$^{\rm a}$
        A.~Ukawa\rlap,$^{\rm a,c}$
        T.~Umeda\rlap,$^{\rm c}$
    and T.~Yoshi\'e$^{\rm a,c}$ }

\begin{document}

\begin{abstract}
We explore the region of small sea quark masses below $m_{PS}/m_V=0.5$
in two-flavor QCD using a mean-field improved clover quark action and 
an RG-improved gauge action at $a \simeq 0.2$ fm on $12^3 \times 24$
and $16^3 \times 24$ lattices. 
We find that instability of the standard BiCGStab algorithm
at small quark masses 
can be mostly removed by the BiCGStab(DS-$L$) algorithm,
which employs $L$-th minimal residual polynomials
with a dynamical selection of $L$. 
We also find singular spikes of $\Delta H$ in the HMC algorithm at moderate 
values of $\Delta\tau$. 
Nature of the spike is studied. 
We also study finite-size effects and chiral properties of meson masses.
\end{abstract}

\maketitle
\thispagestyle{empty}

\section{Introduction}

A rapid increase of the computational cost and instabilities 
in simulation algorithms have so far limited simulations of QCD with dynamical
Wilson-type quarks to quark masses corresponding to 
$m_{PS}/m_{V}$ \simgt 0.6 \cite{kaneko}, to be compared with the physical 
value $m_\pi/m_\rho = 0.18$. 
This limitation of the range of quark masses causes sizable ambiguities 
and systematic errors in the extrapolation to the physical point. 
A related problem is that the mass dependences predicted by chiral perturbation
theory (ChPT) have not been confirmed in full QCD simulations 
\cite{kaneko,Wittig,Hashimoto}.

In this report, we explore the light quark mass region extending our 
previous systematic study of two-flavor QCD \cite{CPPACS-NF2}. 
Through an improvement of simulation algorithms,
sea quark masses corresponding to $m_{PS}/m_{V}=0.6$--0.4 are studied 
on coarse lattices with $a \simeq 0.2$ fm. 


We adopt a renormalization-group improved gauge 
action and a meanfield-improved clover quark action \cite{CPPACS-NF2}. 
At $\beta=1.8$ ($a \simeq 0.2$fm), we employ $12^{3} \times 24$ and 
$16^{3} \times 24$ lattices with the spatial size 2.4 and 3.2 fm. 
Simulation parameters 
and present statistics 
are summarized in Table~\ref{table:simulation_parameters}. 
Measurements are done at every 5 trajectories generated by the 
HMC algorithm.

\begin{table}[b]
\vspace{-4mm}
\caption{Simulation parameters and number of trajectories at $\beta=1.8$.}
\label{table:simulation_parameters}
\leavevmode
\begin{center}
\vspace{-15pt}
\begin{tabular}{lcccc} \hline
\multicolumn{2}{l}{$(m_{PS}/m_{V})_{sea}$}
                              & 0.6     & 0.5     & 0.4   \\ 
$\kappa_{sea}$              & & 0.14585 & 0.14660 & 0.14705 \\ 
\hline
$N_{traj}$ & $12^3 \times 24$ & 3500    & 3500    & 2100    \\ 
           & $16^3 \times 24$ & 1400    & 750     &  --     \\ 
\hline
\end{tabular}
\end{center}
\end{table}


\section{Inversion of the quark matrix}

\begin{figure}[tb]
\vspace{-4mm}
\begin{center}
\leavevmode
\epsfxsize=6.9cm
\epsfbox{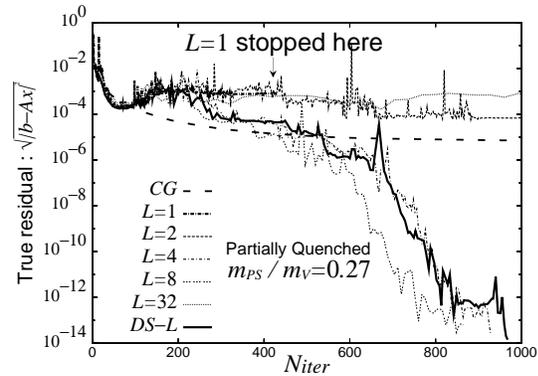}
\end{center}
\vspace{-13mm}
\caption{True residual at $(m_{PS}/m_V)_{val}=0.27$ as a function of 
$L$ and $N_{iter}$ measured on a configuration at $(m_{PS}/m_V)_{sea}=0.6$.
}
\label{fig:residual}
\vspace{-15pt}
\end{figure}

In Ref.~\cite{CPPACS-NF2}, the BiCGStab algorithm is adopted to invert 
the quark matrix. 
At small quark masses, the BiCGStab sometimes fails to converge. 
While the CG algorithm is guaranteed to converge, it is quite time-consuming.
We find that an extension of BiCGStab to $L$-th order minimal residual 
polynomials, the BiCGStab($L$) algorithm \cite{sleijpen}, 
is more stable. 
The conventional BiCGStab corresponds to the case $L=1$.

A larger $L$ is expected to lead to a better convergence.
In practice, however, too large $L$ is time-consuming and also 
frequently introduces other instabilities. 
Figure \ref{fig:residual} shows the $L$-dependence of the residual on a test 
configuration. 
Optimum values of $L$ depend on the simulation parameters. 

To avoid a tuning of $L$ at each simulation point, 
we employ the BiCGStab(DS-$L$) algorithm \cite{miyauchi}.
This is an improvement of BiCGStab($L$) in which an optimum $L$ 
is dynamically selected. 
We find that BiCGStab(DS-$L$) is much more robust than the original BiCGStab 
at small quark masses without much increase of the computer time.

\section{HMC updates}

\begin{figure}[tb]
\leavevmode
\epsfxsize=7.1cm
\epsfbox{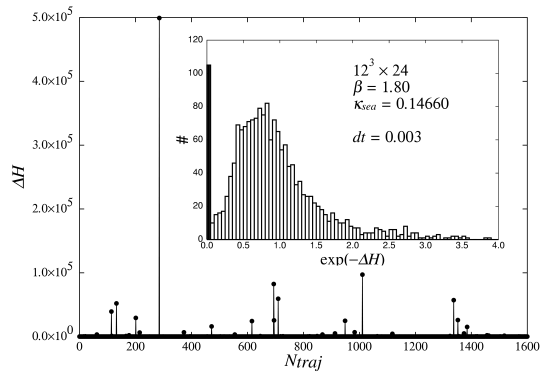}
\vspace{-10mm}
\caption{Time history of $\Delta H$ and the corresponding 
histogram of $\exp(-\Delta H)$.
}
\label{fig:spike_history}
\vspace{-15pt}
\end{figure}

Another difficulty at small quark masses is the appearance of instabilities
in MD evolutions of the HMC algorithm. 
Figure \ref{fig:spike_history} is an example of the time history of 
$\Delta H \equiv H_{trial} - H_{old}$ at $m_{PS}/m_V=0.5$. 
We observe that $\Delta H$ sometimes shows huge values (``spikes'').
A similar phenomenon has been reported in \cite{alpha_1}.
A consequence of the spikes is a distorted distribution of $e^{-\Delta H}$ 
at $e^{-\Delta H}\approx 0$ (see the inset of Fig.~\ref{fig:spike_history}). 
While the trial configurations with large $\Delta H$ are automatically 
rejected by the HMC algorithm, the distortion of the $e^{-\Delta H}$
distribution can introduce additional systematic errors 
(within the statistical errors) 
through the accept/reject process.

We find that the frequency of spikes decreases as we decrease the MD step size 
$dt$.
We have also checked that spikes do not violate the reversibility and 
area-preservation required by the HMC algorithm.

\begin{figure}[tb]
\vspace{-4mm}
\begin{center}
\leavevmode
\epsfxsize=6.9cm
\epsfbox{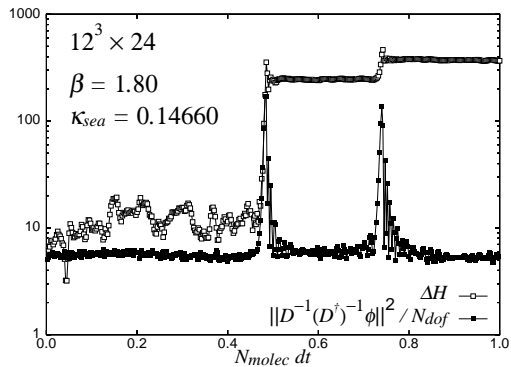}
\end{center}
\vspace{-13mm}
\caption{
$\Delta H$ and the quark force at a spike.
}
\label{fig:spike}
\vspace{-15pt}
\end{figure}

In order to clarify the origin of spikes, we plot 
$\Delta H(p(t),U(t+\frac{1}{2}dt))$ and $||D^{-1}(D^{\dagger})^{-1} \phi||$
for the quark force as functions of the MD time $t$ 
on a configuration with a spike, 
where $p$ is the conjugate momentum of $U$ and 
$\phi$ is the pseudo-fermion field.
From  Fig.~\ref{fig:spike}, we find that $\Delta H$ jumps to a large value 
when the quark force becomes large.
We confirm that the quark force remains small when $dt$ is small enough 
to remove spikes.

A mechanism to induce large $\Delta H$ from large force is 
described in \cite{UKQCD_spike}. 
In a simple harmonic oscillator system with the frequency $\omega$, 
a discrete approximation of the MD evolution becomes unstable for 
$dt > 2/\omega$ and $\Delta H$ diverges exponentially with the MD time. 
This suggests that the discretized MD evolution of HMC with fixed $dt$
may become unstable if an effective $\omega \propto \sqrt{\rm force}$ 
in QCD exceeds a critical value. 
To study the origin of large force, we are currently investigating the 
eigenvalues of the quark matrix.


\section{Meson masses}

\begin{figure}[tb]
\vspace{-4mm}
\begin{center}
\leavevmode
\epsfxsize=6.9cm
\epsfbox{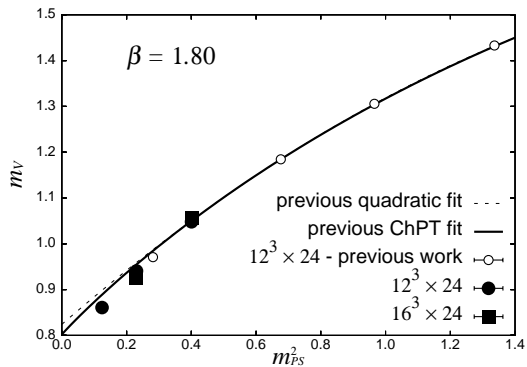}
\end{center}
\vspace{-13mm}
\caption{
Pseudo-scalar and vector meson masses.
}
\label{fig:pseudoscalar_vector_mass}
\vspace{-15pt}
\end{figure}

Our preliminary results for the pseudo-scalar and vector meson masses 
are shown in Fig.~\ref{fig:pseudoscalar_vector_mass}, together with the 
previous results at the same $\beta$ \cite{CPPACS-NF2}. 
We note that the new data points are consistent with the previous 
data in the overlapping region, but show a deviation from the chiral 
extrapolation fit curves of the previous data, at small quark masses. 
The difference amounts to 3\% (2.5$\sigma$) in the chiral limit. 
Comparing the results from $12^{3} \times 24$ and $16^{3} \times 24$ lattices, 
we find that the finite size effects are small at $m_{PS}/m_{V}=0.6$, 
while, at $m_{PS}/m_{V}=0.5$, $m_{V}$ on $16^{3} \times 24$ is slightly 
lower by 2\% (1.5$\sigma$). 
A higher statistics is needed to draw more definite conclusions. 

Using the data shown in Fig.~\ref{fig:pseudoscalar_vector_mass}, 
we test the PCAC relation, 
\begin{equation}
  \frac{m_{PS}^{2}}{2 B_{0} m_{q}}
= 1 + \frac{C y}{N_{f}} \log y
    + \alpha y,
\mbox{\ } y \equiv \frac{m_{PS}^{2}}
                          {(4 \pi f_{PS})^{2}}
\label{eq:PCAC}
\end{equation}
where $B_{0}$ and $\alpha$ are unknown parameters 
and $C=1$ from ChPT. 
Fixing $f_{PS}$ to $f_\pi=93$ MeV 
and adopting the AWI quark mass for $m_q$, 
the data can be well fitted with $\chi^2/df = 0.05$ 
when we treat $C$ as a free parameter. 
However, we obtain $C=0.06(2)$. 
Correspondingly, when we fix $C=1$, the data cannot be described 
by Eq.~(\ref{eq:PCAC}) ($\chi^2/df\approx 40$).  
Therefore, our data down to $m_{PS}/m_{V}=0.4$ do not show the 
logarithmic curvature predicted by ChPT.


\section{Conclusions}

We explored the light quark region of QCD down to $m_{PS}/m_V=0.4$ 
using improved Wilson-type quarks.
We found that the BiCGStab(DS-$L$) algorithm is robust at small quark 
masses. 
Spikes in $\Delta H$ may introduce an additional systematic error
but can be suppressed by setting $dt$ small.
The origin of the huge quark force is still under investigation.
Preliminary analyses on meson masses did not find agreement with 
the logarithmic behavior expected from chiral perturbation theory to one 
loop order. 

\vspace{2mm}

We thank S.\ Itoh and R.\ Frezzotti for useful discussions.
This work is supported in part
by Large Scale Numerical Simulation Project of 
the Science Information Processing Center, University of Tsukuba, 
and by Grants-in-Aid of the Ministry of Education 
(Nos.\ 
11640294, 
12304011, 
12640253, 
12740133,
13135204,
13640259,
13640260,
14046202, 
14740173
). 
VIL is supported by JSPS.




\begin{thebibliography}{99}
\bibitem{kaneko}
For a recent review, see 
T.\ Kaneko, Nucl.\ Phys.\ B (Proc.\ Suppl.) 94 (2002) 133.

\bibitem{Wittig}
H.\ Wittig, these proceedings.

\bibitem{Hashimoto}
S.\ Hashimoto, these proceedings.

\bibitem{CPPACS-NF2}
CP-PACS Collaboration: A.\ Ali Khan {\it et al.},
Phys.\ Rev.\ D65 (2002) 054505.

\bibitem{sleijpen}
G.L.G.\ Sleijpen and D.R.\ Fokkema,
Elec.\ Trans.\ on Numer.\ Anal.\ Vol.1 (1993) 11.

\bibitem{miyauchi}
T.\ Miyauchi {\it et al.}, 
Trans.\ of Japan Soc.\ for Ind.\ and Appl.\ Math.\ 
Vol.11, No.2 (2001) 49.

\bibitem{alpha_1}
K.\ Jansen and R.\ Sommer,
Nucl.\ Phys.\ B530 (1998) 185.

\bibitem{UKQCD_spike}
R.G.\ Edwards {\it et al.},
Nucl.\ Phys.\ B484 (1997) 375;
B.\ Jo\'o {\it et al.},
Phys.\ Rev.\ D62 (2000) 114501.


\end{thebibliography}
\end{document}